\documentclass[twocolumn,aps,prb,showpacs,floatfix]{revtex4}%
\usepackage{amsfonts}
\usepackage{amsmath}
\usepackage{amssymb}
\usepackage{graphicx}
\newcommand{\be}{\begin{eqnarray}}
\newcommand{\en}{\end{eqnarray}}
\newcommand{\beq}{\begin{eqnarray}}
\newcommand{\eeq}{\end{eqnarray}}

\newcommand{\non}{\nonumber}

\newcommand{\para}{\parallel}
\newcommand{\bb}{\mathbf}

\newcommand{\la}{\langle}
\newcommand{\ra}{\rangle}%
\begin{document}
\title{Ferroelectric phase transition, ionicity condensation,
and multicriticality
in  charge transfer organic complexes}
\author{
Jun-ichiro Kishine,$^a$\footnote{Address after September 2003:
\it
Department of Physics,          
   Faculty of Engineering,          
   Kyushu Institute of Technology,
   1-1 Sensuicho, Tobata, Kitakyushu 804-8550, Japan}
Tadeusz Luty,$^b$ and
Kenji Yonemitsu,$^a$
 }
\affiliation{%
$^a$ Department of Theoretical Studies,
Institute for Molecular Science, Okazaki 444-8585, Japan
\\and\\
Department of Functional Molecular Science, Graduate University for Advanced
Studies, Okazaki 444-8585, Japan\\
$^b$Institute of Physical and Theoretical Chemistry
Technical University of Wroclaw
50-370 Wroclaw, Poland
}
\date{\today}

\begin{abstract}
To elucidate a novel pressure-temperature phase diagram of the
quasi-one-dimensional  mixed-stack charge-transfer (CT) complex
TTF-CA,
we study the quasi-one-dimensional spin-1 Blume-Emery-Griffith (BEG) model. 
In addition to the local charge transfer energy and the
inter-stack  polar (dipole-dipole) interaction,
we take account of the inter-stack electrostriction effect.
Using the self-consistent chain-mean-field theory, where
the intra-stack degrees of freedom are exactly treated by the transfer-matrix method,
we reproduce the gas-liquid-solid like phase diagram
corresponding to
the neutral (N), paraelectric ionic (I$_{\rm{para}}$),
and 
 ferroelectric ionic (I$_{\rm{ferro}}$)
phases, respectively. 
We also give an explanation on 
the experimentally observed multicritical behavior 
and concomitant discontinuous inter-stack lattice contraction in  TTF-CA.
\end{abstract}
\maketitle

\section{Introductoin}
In the \lq\lq critical phase control technology,\rq\rq 
condensed molecular materials play quite a promising role, because 
 molecular orbitals and stacking architecture are manipulable
in a desirable way.
To elucidate interrelation of constituent molecular structures and emergence of various
thermodynamic phases such as superconductivity, magnetism, and ferroelectricity is  of great interest there.
A neutral-to-ionic  phase transition (NIT) in 
quasi-one-dimensional 
charge-transfer (CT) complexes  comprising mixed-stack architecture of
 electron donor (D) and accepter (A) molecules\cite{Torrance81a} 
has  played a key role in this field.

In particular, phase control by 
pressure\cite{Cailleau97,Luty02} or laser radiation\cite{Koshihara90,Luty02} in
the tetrathiafulvalene-$p$-chloranil (TTF-CA), that exhibits
the NIT around 80K at ambient 
pressure, has attracted a great deal of interest.
Very recently, Collet at al.,\cite{Collet03} using highly refined time-resolved X-
ray diffraction technique, have reported direct observation of a 
photo-induced paraelectric-to-ferroelectric structural order in the 
crystal.
In the ionic phase, the D$^+$A$^-$ pair forms a dimer
due to the electrostatic instability\cite{Iizuka-Sakano-Toyozawa95} or subsequent
spin-Peierls instability.\cite{Nagaosa88} 
The ionized dimer on the DA chain  carries  
a local electric dipole moment $p$ with opposite
directions depending on the dimerization patterns
\underbar{D$^+$A}$^-$ or \underbar{A$^-$D}$^+$.
Once $p$ acquires a macroscopic mean value $\eta=\langle p\rangle\neq 0$, 
a spontaneous inversion symmetry  breaking (SISB) occurs and
the system undergoes a phase transition to a ferroelectric-ionic (I$_{\rm{ferro}}$) phase.
The ionic phase itself is
simply described by 
ionicity condensation $c=\langle p^2\rangle\sim 1$.
{\it Since $\eta$ is a symmetry-breaking order parameter but $c$ is not, 
we expect that $\eta$ and $c$ play separate roles.}
The
appearance of two distinct order parameters $\eta$ and $c$
is a direct consequence of the
degeneracy of
the two configurations of dimerization pattern, 
IA  
($\cdots$ \underbar{D$^+$A}$^-$\underbar{D$^+$A}$^-$ $\cdots$)
and 
($\cdots$ \underbar{A$^-$D}$^+$\underbar{A$^-$D}$^+$ $\cdots$).

Recently, 
the respective roles of $c$ and $\eta$ have been highlighted in 
both equilibrium\cite{Cailleau97} and
nonequilibrium\cite{Luty02} processes. 
Using the neutron
diffraction along with nuclear-quadrupole-resonance (NQR)
measurements, 
Lem\'ee-Cailleau {\it et al.}\cite{Cailleau97}
 found a novel phase where
{\it the system is ionic but  dipoles  remain disordered}, 
i.e., a paraelectric ionic  (I$_{\rm{para}}$) phase.
They proposed  a pressure-temperature
phase diagram of TTF-CA, where the
N,  I$_{\rm{para}}$, and I$_{\rm{ferro}}$ phases 
are like gas, liquid, and solid phases, respectively.
The ferroelectric order is well signaled  by the appearance of $(0, 2k+1,0)$ Bragg peaks that indicate 
the inversion symmetry breaking.
The  \lq\lq sublimation\rq\rq\,\, line separating 
 the N and I$_{\rm{ferro}}$ phases continues up to a triple point 
 $(P_{t},T_{t})\sim(500{\rm MPa},210{\rm K})$.
Above the triple point, 
in addition to the \lq\lq crystallization (or melting)\rq\rq
\, line, 
there appears a
 \lq\lq condensation\rq\rq \, line separating
the I$_{\rm{para}}$ and I$_{\rm{ferro}}$ 
phases accompanied by
a concomitant
discontinuous change of $c$, ending at a critical 
point $(P_{c},T_{c})\sim(700{\rm MPa},250{\rm K})$. 
The purpose of this paper is to give a qualitative 
understanding of this phase diagram. 
This phase diagram, in addition to the 
phenomenological description offered in ref.\cite{Luty02}.

Since the SISB is prohibited by thermal fluctuations in a purely one-dimensional
stack,
inter-stack  coupling is required to realize the SISB.
In addition, 
the experimental observation strongly indicates that electronic and lattice degrees of freedom 
are
coupled with each other in a unique manner. That is to say, upon crossing the
transition lines in the gas-liquid-solid like phase diagram,\cite{Cailleau97,Luty02}
the unit cell parameter $b$ (for the axis perpendicular to the stack) 
exhibits about 0.5 \%  discontinuous contraction at the condensation transition
but exhibits continuous contraction at the crystallization transition.
On the other hand, the 
unit cell parameter $a$ (for the stacking axis) 
exhibits only continuous contraction 
at the condensation transition
and below it remains almost constant.\cite{Luty02}
Now we are ready to ask the question:
(1) what kind of inter-stack interactions are responsible for the
occurrence of the I$_{\rm{ferro}}$ phase, and
(2) how the lattice anomalies are coupled to the phase transitions ?

As for the first question, 
 Luty {\it et al}.\,\cite{Luty02,Luty96Rev}  stressed that
the inter-stack non-polar coupling\cite{Nagaosa88} alone cannot drive
the ferroelectric ordering and
 the {\it dipolar} coupling plays an essential role.
As for the second question, 
Kawamoto {\it et al}.\cite{Kawamoto01}
took account of the charge distribution on the
atoms inside each molecule
by an ab initio quantum chemical method and
elucidated the importance of  inter-stack Coulomb attraction $\sim-0.14$eV,
which may cause inter-stack electrostriction (Coulomb-lattice coupling).

\section{Quasi-one-dimensional Blume-Emery-Griffith model and interchain mean field theory}
Now we shall set up a model.
The ground-state energy of the
mixed stacks has three minima
as a function of the dimerization displacement, i.e.,
the N and the degenerate IA and IB states.
The three states may be described by the spin-1
Ising variable $p_{i,j}=0,\pm1$ on
the $i$th dimer inside the $j$th stack.\cite{Cailleau97,Luty02,Luty96Rev}
The charge transfer energy ($\Delta$),
the intra- (with subscript $\para$) and inter- (with subscript $\perp$) 
stack 
dipolar ($J$) and non-polar ($K$) interactions,
and the coupling with the
electric field ($E$) are described by
the quasi one-dimensoinal (Q1D) Blume-Emery-Griffith (BEG) model,\cite{BEG71} 
${\cal H}={\cal H}_\para+{\cal H}_\perp$, where
\be
{\cal H}_\para&=&-\sum_{i,j}
\left[J_\para 
p_{i,j}p_{i+1,j}+
K_\para 
p_{i,j}^2p_{i+1,j}^2
\right.
\non\\&&
\left.
\,\,\,\,\,\,\,\,\,\,\,\,\,\,\,\,
\,\,\,\,\,\,\,\,\,\,\,\,\,\,\,\,
\,\,\,\,\,\,\,\,\,\,\,\,\,\,\,\,
-\Delta p_{i,j}^2
-E p_{i,j}
\right],\\
{\cal H}_\perp&=&-\sum_{i,j} 
\left[J_\perp 
p_{i,j}p_{i,j+1}+
K_\perp 
p_{i,j}^2 p_{i,j+1}^2
\right].\label{Q1DBEG}
\en
The intra-stack dipolar interaction $J_\para$ is caused by 
coupling between the charge transfer and the lattice distortion,\cite{Luty96Rev}
while the inter-stack dipolar interaction is regarded as a
direct interaction between the induced dipoles on adjacent stacks.
The intra-stack couplings are much stronger than the
inter-stack couplings,
and the electric dipoles are aligned along the stacks. 
The energy cost to create one D$^+$A$^-$ pair is given 
in the limit of
no molecular overlap by
$
\Delta= I-A-\alpha V,
$
where
$I$ and $A$
denote the donor's ionization energy and the accepter's affinity, respectively,
and $\alpha V$ denotes
the Madelung energy.\cite{MacConnel65}
Generally speaking, increasing pressure decreases
the  
lattice spacing $a$ and consequently increases $V$.
Therefore, $\Delta$ decreases upon applying pressure.

We treat the Hamiltonian (\ref{Q1DBEG}) by using the self-consistent chain mean-field 
theory\cite{SIP75}.
Introducing the thermal averages,
$
\eta=\la p_{i,j}\ra$
and
$
c=\la p_{i,j}^2\ra,
$
we have the
 effective 1D BEG model,
\be
{\cal H}_\para^{\rm eff}
&=&-\sum_{i}
[J_\para p_{i}p_{i+1}+K_\para p_{i}^2p_{i+1}^2
-\tilde\Delta p_{i}^2-\tilde E p_{i}]\non\\
&&
\,\,\,\,\,\,\,\,\,\,\,\,\,\,\,\,
+{z_\perp \over 2} N J_\perp \eta^2
+{z_\perp \over 2}  N K_\perp c^2,
\en
where
$
\tilde\Delta=\Delta-z_\perp K_\perp c$ and
$
\tilde E=E-z_\perp J_\perp \eta,
$
with $z_\perp=2$ being the inter-stack coordination number.
Treating 
$
{\cal H}_\para^{\rm eff}
$
exactly by the transfer matrix method, 
we obtain
the free energy per site,
\be
f_{\rm{BEG}}(\eta,c,T)=-T\ln \lambda(\tilde\Delta,\tilde E,T)
+ J_\perp \eta^2+ K_\perp c^2,\label{fene}
\en
where
$\lambda(\tilde\Delta,\tilde E,T)$
is the maximum eigenvalue of the transfer matrix for ${\cal H}_\para^{\rm eff}$, given by
\be
\bb{T}\!\!=\!\!
\left(
\begin{array}{ccc} 
e^{\beta(J_\para+K_\para-
\tilde
\Delta-
\tilde E)}
\!\!
&e^{-\beta(\tilde\Delta+\tilde E)/2}\!\!
&e^{\beta(-J_\para+K_\para-\tilde\Delta)}\!\!\cr 
e^{-\beta(\tilde\Delta+\tilde E)/2} \!\!& 1\!\! & e^{-\beta(\tilde\Delta-\tilde E)/2}\!\!\cr 
e^{\beta(-J_\para+K_\para-\tilde\Delta)}\!\!&e^{-\beta(\tilde\Delta-\tilde E)/2}\!\!
&e^{\beta(J_\para+K_\para-\tilde\Delta+\tilde E)}\!\!
\end{array}
\right),
\en
with
$\beta=1/T$.
Possible phase diagrams of the BEG model have been extensively studied
through 
mean-field theories,\cite{BEG71,Hoston-Berker91} renormalization-group,\cite{Berker-Wortis76} and
 transfer-matrix methods.\cite{Albayrak-Keskin00}
For the parameter regions relevant to the present case, 
$J_\para$,
$J_\perp$, 
$K_\para$, 
$K_\perp$, and $\Delta$ are all positive, 
so that a solid-liquid-gas type phase diagram with proper
slopes of transition lines
is not obtained.

\section{Interchain electrostriction}

Then, we consider the inter-stack lattice degrees of  freedom 
that have not explicitly been  taken into account
in (\ref{Q1DBEG}).
It is well known that an electrostriction effect potentially
converts a continuous transition to a discontinuous one,
since this 
gives rise to an additional negative free-energy
 term that contains the forth power of the relevant order parameter.\cite{BasicNotion}
 In the present case, 
we phenemenologically introduce 
 an additional free energy,
\be
f_{\rm{elst}}(c,y)
=-{c^2\over b_0+y}+{1\over 2}k y^2,
\en
where the first and second terms represent Coulomb attraction between the
nearest neighbor stacks\cite{Kawamoto01} 
and the elastic energy for the distortion in the inter-stack
direction. 
Note that $f_{\rm{elst}}(c,0)$ has already been absorbed into $K_\perp$.
The lattice constant without 
distortion is $b_0$, 
and $y$ denotes the distortion.
By minimizing $f_{\rm{elst}}(c,y)$ with respect to $y$, we obtain
the optimized lattice constant,
\be
b(T)=b_0+y(T)\sim
b_0-2{\varepsilon}_{\rm{elst}}
b_0^2 c^2,\label{latticeparam}
\en
where ${\varepsilon}_{\rm{elst}}=1/(2kb_0^4)$ is a small 
constant. 
We thus have the energy gain due to the lattice distortion,
\be
f_{\rm{elst}}(c,T)
\sim -{\varepsilon}_{\rm{elst}}c^4.
\en
Now, solving the self-consistent equations is reduced to searching
$(c,\eta)$ that gives the absolute minimum of the total free energy,
$
f(\eta,c,T)=f_{\rm{BEG}}(\eta,c,T)+f_{\rm{elst}}(c,T).
$
The ionic phase is characterized by  the ionicity condensation $c=1$,
while the  ferroelectric phase is characterized by $\eta\neq0$. 

Note that, in the present scheme, 
any phase with $c\neq 1$ is regarded as \lq\lq neutral\rq\rq\,\,
and that the neutral phase has always $\eta=0$.
In the BEG model, because of three states $p_{i,j}=0,\pm1$, 
 $c$ approaches the universal constant $c=2/3$ 
 in the high temperature limit, where the entropy term dominates the internal 
 energy term.
 Therefore, in the parameter region where $c$ continuously increases upon 
 decreasing temperature, we have $2/3<c\leq 1$.
In the experiments,\cite{Torrance81a} 
the ionicity continuously increases upon 
 decreasing temperature and jumps from $c\sim 0.3$ to $c\sim 0.6$ at
the NIT.
Thus, concerning the quantitative magnitude of the ionicity,
there arises a difference between the experimental result and
the present analysis. 
This apparent difference comes from the fact that
we mapped the  intra-stack
CT transfer and the DA dimerization 
onto the simple spin-1 Ising variables.
Therefore, we should regard the difference 
as an  artifact of the classical BEG model.

\section{Numerical results and phase diagram}

\begin{figure}[h]
\includegraphics[width=9cm]{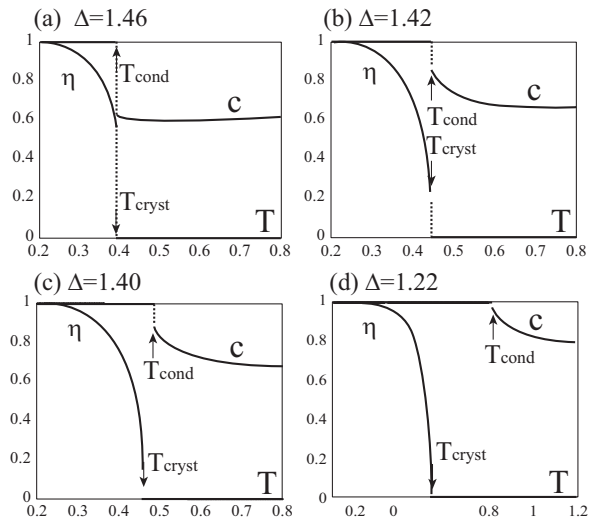}
\caption{ 
Temperature dependence of $c$ and $\eta$
for various magnitudes of $\Delta$ with
$K_\para=0.4$,
$K_\perp=0.06$,
$J_\perp=0.03$,
and 
$
{\varepsilon}_{\rm{elst}}=0.0095.
$
The condensation and crystallization temperature, 
$T_{\rm{cond}}$ and $T_{\rm{cryst}}$, respectively,
are indicated.
Locations of the $\Delta$ values in (a)-(d) 
are indicated in the phase diagram of Fig.~\ref{pd}.
}
\label{ceta}
\end{figure}

From now on, we set $J_\para=1$ as an energy unit and
the electric field $E$ is set to be zero.
In Fig.~\ref{ceta},
we show the temperature dependence of $c$ and $\eta$
for various magnitudes of $\Delta$ with
$K_\para=0.4$,
$K_\perp=0.06$,
$J_\perp=0.03$,
and 
$
{\varepsilon}_{\rm{elst}}=0.0095.
$
We introduce the condensation temperature $T_{\rm{cond}}$
and the crystallization temperature $T_{\rm{cryst}}$.
The ionicity jumps into $c=1$ at $T_{\rm{cond}}$, while
the ferroelectric order parameter 
acquires a finite magnitude $\eta\neq 0$ at $T_{\rm{cryst}}$.
The ground state becomes ionic for $\Delta<1.49$.
Both $c$ and $\eta$ exhibits a discontinuous change at the same  transition temperature
(the sublimation temperature)
for $1.42<\Delta<1.49$.
That is to say, 
$
T_{\rm{cond}}
=
T_{\rm{cryst}}
$.
For $\Delta<1.42$, 
there appears a region, $T_{\rm{cryst}}<T<T_{\rm{cond}}$, 
where the system is  ionic but still paraelectric.
This region is identified with the $I_{\rm{para}}$ phase
that is observed in TTF-CA under pressure.
The point $(\Delta_t=1.42,T_t=0.45)$ is identified with the {\it triple point}
(indicated by \lq\lq TP\rq\rq\,\, in Fig.~\ref{pd}).
For  $\Delta<1.42$,    $c$ still exhibits a
discontinuous change at $T_{\rm{cond}}$, but
$\eta$ continuosly evolves at $T_{\rm{cryst}}$, 
as shown in Fig.~\ref{ceta}(c).
The discontinuity jump of $c$ becomes smaller
and seems to vanish,
as $\Delta$ decreases,
as shown in Fig.~\ref{ceta}(d).
We stress that
this discontinuity is a direct consequence of the weak but finite
electrostriction effect.
Without the electrostriction, as $\Delta$ decreases, 
$T_{\rm{cond}}$ and $T_{\rm{cryst}}$ continue to coincide with each other, and the transition
simply changes from  discontinuous 
 to  continuous  at some critical value of 
$\Delta$.\cite{BEG71,Hoston-Berker91,Berker-Wortis76,Albayrak-Keskin00}

\begin{figure}
\includegraphics[width=8cm]{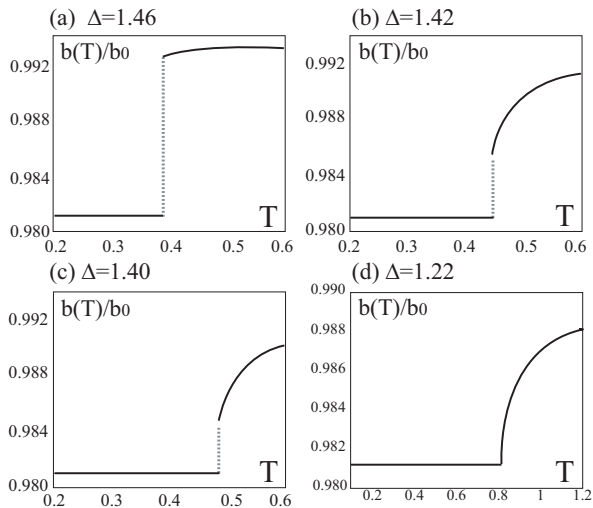}
\caption{ 
Temperature dependence of the  unit cell parameter $b$, $b(T)$.
Locations of the $\Delta$ values in (a)-(d) 
are indicated in the phase diagram of Fig.~\ref{pd}.
}\label{latt}
\end{figure}

As clearly seen from (\ref{latticeparam}),
about 0.5\% discontinuous contraction of the
inter-stack lattice
constant (unit cell parameter $b$)  
is accompanied by
the discontinuous jump of the ionicity. 
In Fig.~\ref{latt},
setting $b_0$ as a length unit,
we show the temperature dependence of the  unit cell parameter $b$
given by (\ref{latticeparam}),
using the same parameter set as that in Fig.~\ref{ceta}. 
Although the  magnitude of the discontinuous contraction depends on the parameter choice of
${\varepsilon}_{\rm{elst}}$ and $b_0$,  
the qualitative nature ($b$ jumps at $T_{\rm{comd}}$) does not change.

\begin{figure}
\includegraphics[width=4.5cm]{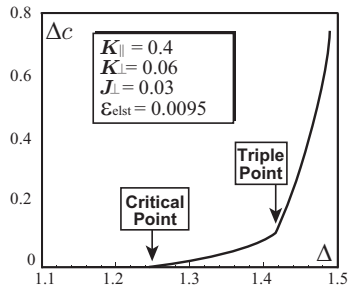}
\caption{ 
Temperature dependence of the discontinuity
of the ionicity, $\Delta c$.
}
\label{discon}
\end{figure}

To  see the discontinuity of the ionicity more closely, 
we show in  Fig.~\ref{discon}
the $\Delta$ dependence of the discontinuity
at the condensation temperature, $\Delta c$.
It is clearly seen that $\Delta c$ 
decreases as $\Delta$ decreases and
eventually reaches zero at $\Delta=1.25$.
For $\Delta<1.25$,
 the condensation occurs without  ionicity jump.
Then, the lattice contraction at $T_{\rm{cond}}$ also becomes continuous.
Therefore,  $\Delta=1.25$ with the corresponding $T_{\rm{cond}}=0.76$
is idendentified with a {\it critical point} (indicated by \lq\lq CP\rq\rq\,\, in Fig.~\ref{pd}).
This result is well consistent with the experimental fact
that the ionicity jump finishes at the critical point.\cite{Cailleau97}

\begin{figure}
\includegraphics[width=8cm]{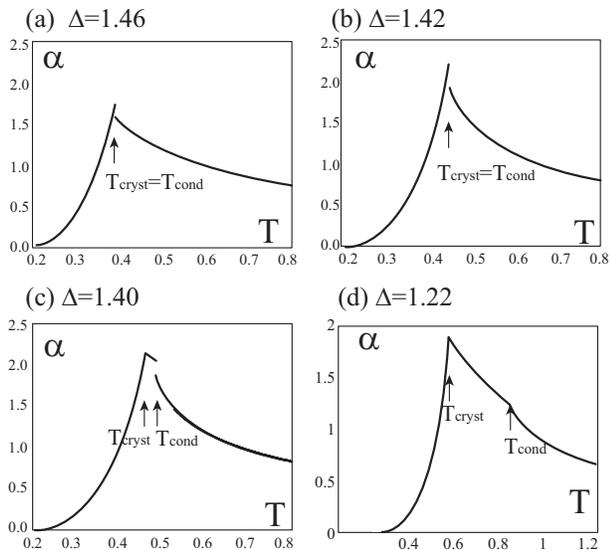}
\caption{ 
Temperature dependence of the
polarizability, $\alpha$. 
Locations of the $\Delta$ values in (a)-(d) 
are indicated in the phase diagram of Fig.~\ref{pd}.}
\label{de}
\end{figure}

\begin{figure}[h]
\includegraphics[width=8.5cm]{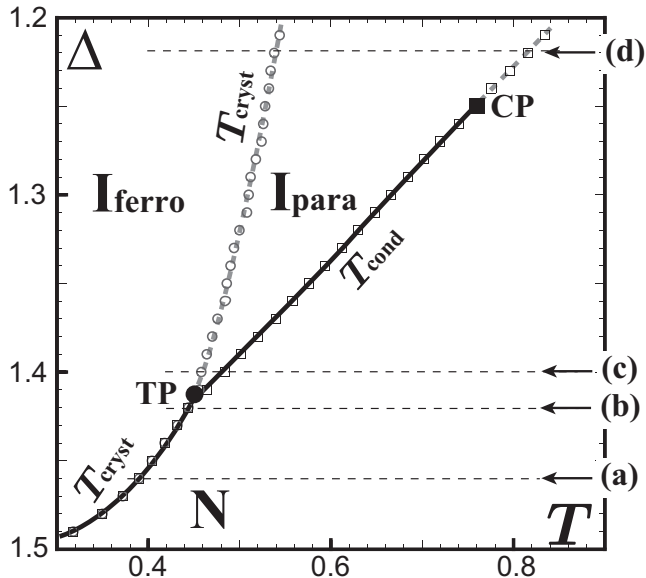}
\caption{ 
Phase diagram
for 
$K_\para=0.4$,
$K_\perp=0.06$,
$J_\perp=0.03$,
and 
$
{\varepsilon}_{\rm{elst}}=0.0095.$
The solid  and  dashed lines
represent discontinuous and continuous transitions,
respectively.
{\bf TP}
and
{\bf CP}
represent the triple point and 
the critical  point, respectively.
Locations of the $\Delta$ values used in 
(a)-(d) of Figs.~\ref{ceta}, \ref{latt}, and \ref{de}
are indicated by the horizontal arrows.
}
\label{pd}
\end{figure}

The dielectric constant is
given by $\varepsilon=1+4\pi\alpha$,
where the uniform
polarizability is  
$
\alpha={1\over T}
\sum_{i,j}
\sum_{l,m}
\left[
\langle
p_{i,j}p_{l,m}
\rangle
-
\langle
p_{i,j}
\rangle
\langle
p_{l,m}
\rangle
\right]
=
(c-\eta^2)/T.
$
In Fig.~\ref{de},
we show the temperature dependence of $\alpha$ for
various magnitudes of $\Delta$.
It is seen that
along the N-I$_{\rm{ferro}}$
boundary,  the polarizability exhibits
a sharp single cusp at $T_{\rm{cond}}=T_{\rm{cryst}}$.
For  $\Delta_c<\Delta<\Delta_{t}$,
a discontinuous jump occcurs at $T=T_{\rm{cond}}$
and a cusp at $T=T_{\rm{cryst}}$.
The discontinuity at 
$T=T_{\rm{cond}}$
finishes at $\Delta=\Delta_c$.

In Fig.~\ref{pd}, we show the phase diagram of the system
for 
$K_\para=0.4$,
$K_\perp=0.06$,
$J_\perp=0.03$,
and 
$
{\varepsilon}_{\rm{elst}}=0.0095.
$
Regarding the decreasing $\Delta$ as 
increasing pressure,
this phase diagram is 
consistent with
the experimentally found,  pressure-temperature
phase diagram of TTF-CA.\cite{Cailleau97}
The tiriple point, the
critical  point, and the observed inter-stack lattice contraction
are reproduced.
For simplicity, we here ignored the change of $\Delta$ due to  thermal
lattice contraction.
Exactly speaking, to convert our $\Delta$-$T$ phase diagram to
a $P$-$T$ diagram, we need to 
take account of the temperature dependence of 
$\Delta$, $\Delta(T)$.
By appropriately treating $\Delta(T)$, 
we may obtain the corresponding $P-T$ phase diagram
satisfying
the Clausius-Clapeyron relation.
We stress that,
even when we take this simple view, a qualitative nature of the phase diagram is
not changed. 
Identifying 
the triple point $(\Delta_t,T_t)=(1.42,0.45)$
with the experimentally obtained one
$(P_{t},T_{t})\sim(500{\rm MPa},210{\rm K})$,
we see that our parameter choice here corresponds to
$K_\perp=28$K and $J_\perp=14$K.

Lajzerowicz and Sivardi\'ere\cite{Lajzerowicz75}  extensively 
developed a mean-field analysis of  the BEG model
and obtained liquid-gas-solid like phase diagrams on the $P$-$T$ plane.
However, they considered  a lattice gas analogue of a simple fluid, where
the physical pressure of the lattice gas is simply given by $-f$, with $f$
being the Helmholtz free energy  per volume.
In the present context,  the pressure of the spin system 
has no  physical meaning
and the phase diagram obtained by Lajzerowicz and Sivardi\'ere cannot be applied to TTF-CA.

\section{Concluding remarks}

In this paper, we showed that 
the  inter-stack electrostriction 
causes the $-c^4$ term with a
small coefficient 
and makes the phase diagram rich.
We thus conclude that 
the inter-stack polar interaction together with
the inter-stack electrostriction
drives the discontinuous inter-stack lattice contraction and
the multicritical behavior observed in  TTF-CA.\cite{Cailleau97,Luty02}
Our considerations are limited to classical models, i.e.,
we considered  classical {\it effective} models, where
all the microscopic (electronic or phonon) 
degrees of freedom are implicitly integrated out.
To obtain the truly microscopic mechanism 
 behind the multicriticality, we need to get back to such
a microscopic Hamiltonian as an extended Peierls-Hubbard model.\cite{Miyashita03}
We would keep this issue for  future study.


This work was partly supported by a  Grant-in-Aid for Scientific Research (C)
from Japan Society for the Promotion of Science.
TL wishes to acknowledge hospitality at the Institute for Molecular Sciences, Okazaki, and thank
colleauges for creating a pleasant and stimulating environment there.

\bigskip

\end{document}